\documentclass[]{ptephy} % oka

\preprintnumber{XXXX-XXXX} %%% %%% Insert preprint number here

\begin{document}

\title{Search for solar Kaluza-Klein axion by annual modulation with the XMASS-I detector}

%%%% To generate auto affiliation numbers please use \author{}\affil{} command

\newcommand{\ICRR}{1}
\newcommand{\IBS}{2}
\newcommand{\ISEE}{3}
\newcommand{\Tokushima}{4}
\newcommand{\IPMU}{5}
\newcommand{\KMI}{6}
\newcommand{\Kobe}{7}
\newcommand{\KRISS}{8}
\newcommand{\Miyagi}{9}
\newcommand{\Tokai}{10}
\newcommand{\YNU}{11}
\newcommand{\Tohokunow}{\dagger}
\newcommand{\Tokushimanow}{\ddagger}

\author{\name{{\bf XMASS Collaboration}}{\ast}\\
\name{N.~Oka}{\Kobe},
\name{K.~Abe}{\ICRR,\IPMU},
\name{K.~Hiraide}{\ICRR,\IPMU},
\name{K.~Ichimura}{\ICRR,\IPMU},
\name{Y.~Kishimoto}{\ICRR,\IPMU},
\name{K.~Kobayashi}{\ICRR,\IPMU},
\name{M.~Kobayashi}{\ICRR},
\name{S.~Moriyama}{\ICRR,\IPMU},
\name{M.~Nakahata}{\ICRR,\IPMU},
\name{T.~Norita}{\ICRR},
\name{H.~Ogawa}{\ICRR,\IPMU},
\name{K.~Sato}{\ICRR},
\name{H.~Sekiya}{\ICRR,\IPMU},
\name{O.~Takachio}{\ICRR},
\name{A.~Takeda}{\ICRR,\IPMU},
\name{S.~Tasaka}{\ICRR},
\name{M.~Yamashita}{\ICRR,\IPMU},
\name{B.~S.~Yang}{\ICRR,\IPMU},
\name{N.~Y.~Kim}{\IBS},
\name{Y.~D.~Kim}{\IBS},
\name{Y.~Itow}{\ISEE,\KMI},
\name{K.~Kanzawa}{\ISEE},
\name{R.~Kegasa}{\ISEE},
\name{K.~Masuda}{\ISEE},
\name{H.~Takiya}{\ISEE},
\name{K.~Fushimi}{\Tokushima,}\thanks{Now at Department of Physics, Tokushima University, 2-1 Minami Josanjimacho Tokushima city, Tokushima, 770-8506, Japan},
\name{G.~Kanzaki}{\Tokushima},
\name{K.~Martens}{\IPMU},
\name{Y.~Suzuki}{\IPMU},
\name{B.~D.~Xu}{\IPMU},
\name{R.~Fujita}{\Kobe},
\name{K.~Hosokawa}{\Kobe,}\thanks{Now at Research Center for Neutrino Science, Tohoku University, Sendai, Miyagi 980-8578, Japan},
\name{K.~Miuchi}{\Kobe},
\name{Y.~Takeuchi}{\Kobe,\IPMU},
\name{Y.~H.~Kim}{\KRISS,\IBS},
\name{K.~B.~Lee}{\KRISS},
\name{M.~K.~Lee}{\KRISS},
\name{Y.~Fukuda}{\Miyagi},
\name{M.~Miyasaka}{\Tokai},
\name{K.~Nishijima}{\Tokai},
\name{S.~Nakamura}{\YNU}
}

\address{
\affil{\ICRR}{Kamioka Observatory, Institute for Cosmic Ray Research, the University of Tokyo, Higashi-Mozumi, Kamioka, Hida, Gifu 506-1205, Japan}
\affil{\IBS}{Center of Underground Physics, Institute for Basic Science, 70 Yuseong-daero 1689-gil, Yuseong-gu, Daejeon 305-811, South Korea}
\affil{\ISEE}{Institute for Space-Earth Environmental Research, Nagoya University, Nagoya, Aichi 464-8601, Japan}
\affil{\Tokushima}{Institute of Socio-Arts and Sciences, The University of Tokushima, 1-1 Minamijosanjimacho Tokushima city, Tokushima, 770-8502, Japan}
\affil{\IPMU}{Kavli Institute for the Physics and Mathematics of the Universe (WPI), the University of Tokyo, Kashiwa, Chiba 277-8582, Japan}
\affil{\KMI}{Kobayashi-Maskawa Institute for the Origin of Particles and the Universe, Nagoya University, Furo-cho, Chikusa-ku, Nagoya, Aichi 464-8602, Japan}
\affil{\Kobe}{Department of Physics, Kobe University, Kobe, Hyogo 657-8501, Japan}
\affil{\KRISS}{Korea Research Institute of Standards and Science, Daejeon 305-340, South Korea}
\affil{\Miyagi}{Department of Physics, Miyagi University of Education, Sendai, Miyagi 980-0845, Japan}
\affil{\Tokai}{Department of Physics, Tokai University, Hiratsuka, Kanagawa 259-1292, Japan}
\affil{\YNU}{Department of Physics, Faculty of Engineering, Yokohama National University, Yokohama, Kanagawa 240-8501, Japan}
\email{xmass.publications9@km.icrr.u-tokyo.ac.jp}}
%%% To include the collaborator name... Please use the command "\collaborator"
%%% For example: \collaborator{ATLAS Collaboration}

\begin{abstract}%
In theories with large extra dimensions beyond the standard 4-dimensional spacetime, axions could propagate in such extra dimensions, and acquire Kaluza-Klein (KK) excitations.
These KK axions are produced in the Sun and could solve the unexplained heating of the solar corona.
While most of the solar KK axions escape from the solar system, a small fraction is gravitationally trapped in orbits around the Sun.
They would decay into two photons inside a terrestrial detector. 
The event rate is expected to modulate annually depending on the distance from the Sun.
We have searched for the annual modulation signature using $832\times 359$ kg$\cdot$days of XMASS-I data.
No significant event rate modulation is found, and hence we set the first experimental constraint on the KK axion-photon coupling of $4.8 \times 10^{-12}\, \mathrm{GeV}^{-1}$ at 90\% confidence level for a KK axion number density of $\bar{n}_\mathrm{a} = 4.07 \times 10^{13}\, \mathrm{m}^{-3}$, the total number of extra dimensions $n = 2$, and the number of extra dimensions $\delta = 2$ that axions can propagate in.
\end{abstract}

\subjectindex{C14, C15, C43}
%C43: underground physics
%C14 Extra dimensions (experiment)
%C15 Axions (experiment)
\maketitle

\section{Introduction}
In quantum chromodynamics (QCD) there is a problem known as the strong CP problem.
The QCD Lagrangian has a CP-violating term, but this violation has not been observed experimentally.
Peccei and Quinn introduced a new global symmetry $U(1)_\mathrm{PQ}$ to solve this problem \cite{Peccei1977}.
This newly introduced symmetry spontaneously breaks at an energy scale of $f_\mathrm{PQ}$, which predicts the existence of a pseudo-Goldstone boson, namely the axion. 
From the experimental results, $f_\mathrm{PQ}$ is found to be far larger than the electro-weak scale \cite{ADD1999}.
The mass of the axion is characterized by $f_\mathrm{PQ}$ and calculated \textcolor{black}{to be}:
\begin{align}\label{eq:mass}
m_\mathrm{a} = 6 \times 10^{15}\, \mathrm{eV}^2 / f_\mathrm{PQ}.
\end{align}
\par
On the other hand, large extra dimensions are also proposed to solve the gauge hierarchy problem \cite{ADD1998, AADD1998}.
Motions of a particle in the extra dimensions could be observed as new particles with heavier masses in the standard 4-dimensional spacetime.
These particles are called Kaluza-Klein (KK) particles.
Although the PQ model and large extra dimensions are independently motivated, studies to link them emerged in the late 1990s.
It was pointed out that large extra dimensions can explain the largeness of $f_\mathrm{PQ}$ \cite{ADD1999}.
Following this possible close connection between the PQ model and extra dimensions, studies of axions in extra dimensions, or KK axions, started \cite{Dienes2000, Chang2000}.  
\par
This opened a possibility of the detection of KK axions produced in the Sun, also called the solar KK axion, in terrestrial detectors \cite{DiLella2000}.
It was also pointed out as an independent motivation for the existence of KK axions that solar KK axions may solve some astrophysical observational problems such as solar coronal heating and X-rays from the dark side of the Moon \cite{DiLella2003}.
In this scenario, a small fraction of the solar KK axions is gravitationally trapped in orbits around the Sun and accumulated over the age of the Sun.
These KK axions then decay into two photons, which become a source of the lunar X-rays.
The X-rays from KK axions trapped near the solar surface on the other hand may explain the observed coronal heating.
It is argued that a gas time projection chamber (TPC) for direct dark matter searches in an underground laboratory is one of the most suitable instruments to detect the decay of KK axions because of its large volume ($\sim 1\, \mathrm{m}^3$) and the ability to distinguish two photons by recording their tracks \cite{Morgan2005}.
No experimental result on solar KK axions has, however, been reported to date. 
\par
XMASS-I is a large-volume liquid xenon scintillation detector designed for multiple physics targets \cite{Detector}.
A variety of searches had been performed using both nuclear and electron channels \cite{LightWIMP, SolarAxion, Inelastic, SuperWIMP, DEC}.
Thanks to the high photoelectron (PE) yield (about $15\, \mathrm{PE/keV}$), the energy threshold is low enough to search for solar KK axions.
XMASS-I data accumulated over more than one year was used for a dark matter direct search in an annual modulation analysis \cite{Modulation}.
Annual modulation is expected also for the solar KK axion signal due to the seasonal change in the distance between the Sun and the Earth.
Annual modulation searches have the advantage of being robust against most backgrounds.
In this study, we conducted the first direct search for solar KK axions by exploiting this advantage.
\section{The XMASS-I detector}
The XMASS-I detector is located 1000~m (2700~m water equivalent) underground at the Kamioka Observatory in Japan.
Its detailed design and performance were described elsewhere \cite{Detector}.
The inner detector (ID) consists of 832~kg of liquid xenon surrounded by 642 inward-looking 2-inch  photomultiplier tubes (PMTs) \textcolor{black}{arranged on a pentakis-dodecahedron-shaped copper holder.
The photocathode coverage of the detector's inner surface is about 62\%.
}
The sensitive volume of the liquid xenon is $0.288~\mathrm{m}^3$.
This ID is surrounded by an outer detector (OD), which is a water tank 10.5~m in height and 10~m in diameter.
The OD has 72 20-inch PMTs and works as a Cherenkov veto counter and a passive shield. 
The threshold for the ID PMTs is equivalent to 0.2 PE.
A trigger is issued when four or more ID PMTs have signals exceeding the threshold within 200~ns.
The signals from the inner detector PMTs are recorded by CAEN V1751 waveform digitizers \textcolor{black}{with a 1~GHz sampling rate}.

\par
Radioactive sources can be inserted into the ID for calibration.
The energy scale is obtained from various sources with energies between 5.9~keV and 122~keV measured, specifically with $^{55}$Fe, $^{57}$Co, $^{109}$Cd, and $^{241}$Am sources \cite{Source}, and its time variation is traced by weekly $^{57}$Co calibrations.
In this paper, the $\mathrm{keV_{ee}}$ energy scale is used, which reflects the electron equivalent energy.
\par
A detailed Monte Carlo (MC) simulation based on Geant4 \cite{Geant4} has been developed.
It simulates particle tracks, the scintillation process, photon tracking, PMT response, and the readout electronics and is used to calculate the expected signal.
The non-linearity of the energy scale is taken into account in the MC using the calibration results and the non-linearity model from Doke et al. \cite{Doke2001}.
The energy scale below 5.9~keV is estimated by an extrapolation based on Doke model.
%A 15\% energy scale difference from the Noble Element Simulation Technique (NEST) \cite{NEST} at the threshold energy of 1.1 keVee ($\sim$ 8 photoelectrons) was found.
%
%
\section{Expected signal}
Solar KK axions would be produced \textcolor{black}{thermally} in the Sun via the Primakoff effect ($\gamma +  Ze \to Ze + \mathrm{a}$, where $Ze$ represents the charge of a nucleus) and a photon coalescence mechanism ({$\gamma + \gamma \to \mathrm{a}$).
Since there are an infinite number of KK excitations, KK axions with a spectrum of masses would be produced according to the temperature of the Sun.
Most of the produced KK axions escape from the solar system, but a small fraction is trapped by the gravity of the Sun.
Figure~\ref{fig:density} shows the expected number density of trapped KK axions against the distance from the Sun.
This distribution is taken from  L.~Di~Lella and K.~Zioutas \cite{DiLella2003} who calculated up to $200R_\odot$ and it is fitted well by $r^{-4}$.
Here, $R_\odot$ represents the radius of the Sun.
The KK axion-photon coupling $g_{a\gamma\gamma} = 9.2 \times 10^{-14}\, \mathrm{GeV}^{-1}$ \textcolor{black}{is} fixed \textcolor{black}{so that axion decay can explain} the X-ray surface brightness of the quiet Sun.
Also, the total number of extra dimensions $n = 2$, the number of extra dimensions $\delta = 2$ that axions can propagate in, and the fundamental scale $M_\mathrm{F} = 100\, \mathrm{TeV}$ are assumed.
\begin{figure}\centering
\includegraphics[width=10cm]{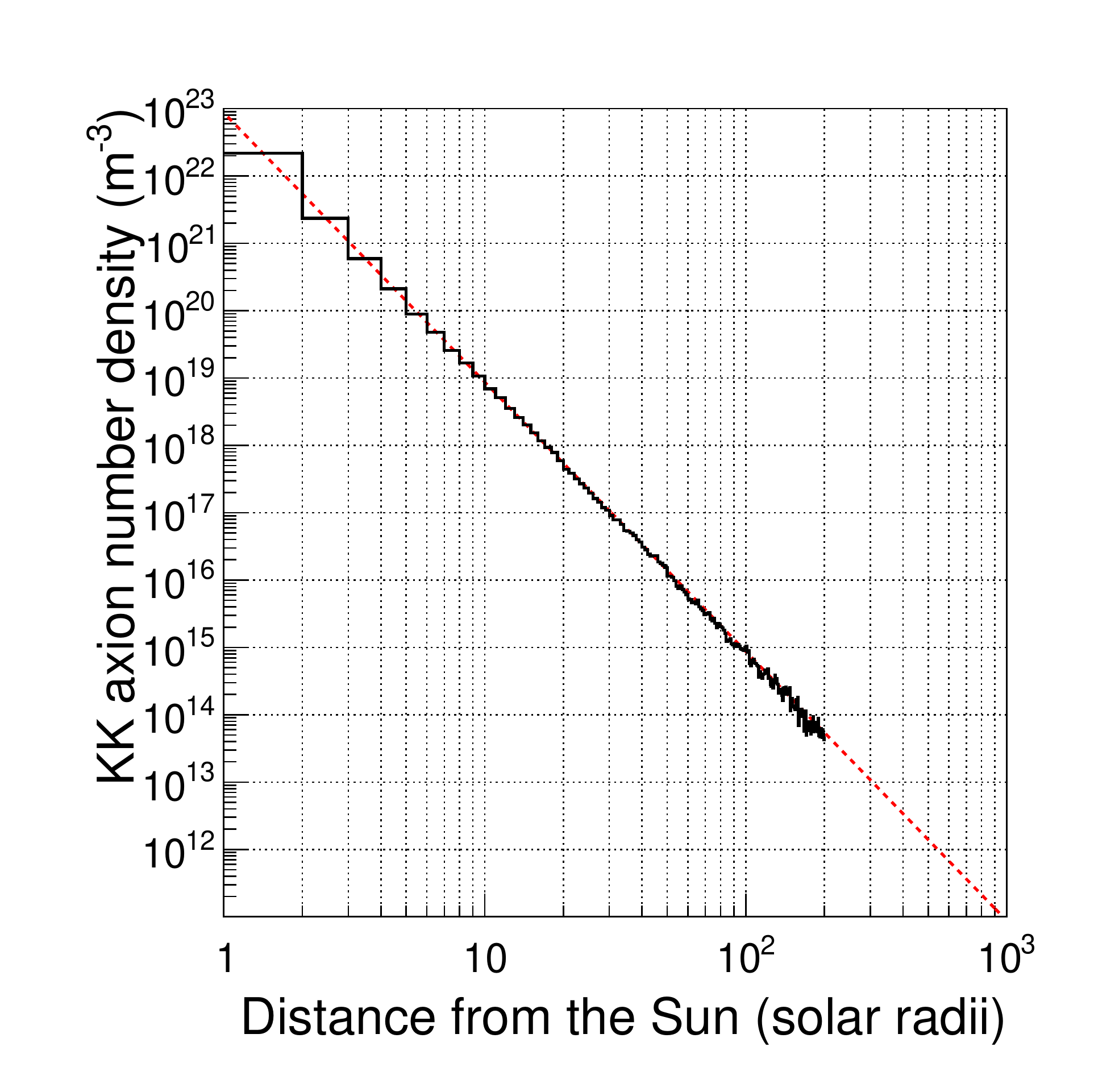}
\caption{Number density of trapped KK axions against the distance from the Sun. The black solid histogram shows the simulated distribution taken from \cite{DiLella2003} with $g_{a\gamma\gamma} = 9.2 \times 10^{-14}\, \mathrm{GeV}^{-1}$. The red dotted line shows the fitted $r^{-4}$ curve. The unit of distance is the radius of the Sun (695,700 km). Using this unit, the distance between the Earth and the Sun is $211.4R_\odot$ at perihelion and $218.6R_\odot$ at aphelion, respectively.}
\label{fig:density}
\end{figure}
The mass dependence of the production rate and the rate of KK axion trapping are also calculated in Ref. \cite{DiLella2003}.
Since the number of trapped KK axions produced by \textcolor{black}{the} Primakoff effect is three orders of magnitude smaller than those produced by the photon coalescence mechanism, we concentrate on the contribution from photon coalescence.
Since the trapped KK axions are non-relativistic, the two photons from their decay have an energy of $m_\textrm{a}/2$ each.
Heavier KK axions are likely to be trapped and would have a shorter life time.
\textcolor{black}{From} the simulation in Ref.~\cite{DiLella2003}, which considers the mass dependencies of the trapping rate and the decay rate, the peak of the decay spectrum of the trapped KK axion near the Earth is about $9~\mathrm{keV}$.
\par
The expected annual modulation signal is calculated as follows: 
First, the distance between the Earth and the Sun as a function of time $r(t)$ is denoted as
\begin{align}\label{eq:r}
r(t) = a \left(1-e\cos \frac{2\pi(t-t_0)}{T}\right),
\end{align}
where $a = 1.496 \times 10^8\, \mathrm{km} = 215.0 R_\odot$ and $e = 0.0167$ are the semi-major axis and the eccentricity of the Earth's orbit, respectively.
$t$ is a date in one year, and $T$ represents one year.
$t_0$ represents the date when the Earth is at perihelion.
\par
Then, since the  number density of the KK axion ($n_\mathrm{a}$) can be fitted with \textcolor{black}{a} $r^{-4}$ curve, the time dependence of the number density $n_\mathrm{a}(t)$ is written as
\begin{align}\label{eq:na}
n_\mathrm{a}(t) &= \bar{n}_\mathrm{a} \left(1-e\cos \frac{2\pi(t-t_0)}{T}\right)^{-4} \notag \\
&\approx \bar{n}_\mathrm{a} \left[1 + 4e \left(\cos \frac{2\pi(t-t_0)}{T} +\frac{5}{2} e\cos^2 \frac{2\pi(t-t_0)}{T} \right)\right].
\end{align}
Here, $\bar{n}_\mathrm{a}$ is the KK axion number density when $r(t) = a$.
From Eq.~(\ref{eq:na}) and Fig.~\ref{fig:density}, the number density of trapped KK axions on the Earth is calculated as  $4.36 \times 10 ^{13}\, \mathrm{m}^{-3}$ at perihelion and $3.81 \times 10 ^{13}\, \mathrm{m}^{-3}$ at aphelion.
\par
The expected KK axion decay rate $R$ is proportional to the square of the KK axion-photon coupling $g_{\mathrm{a}\gamma\gamma}$ and the number density \cite{Morgan2005}:
\begin{align}\label{eq:decay}
R = \left(2.5\times 10^{11}\, \mathrm{m^{-3}day^{-1}}\right) \left( \frac{g_{\mathrm{a}\gamma\gamma}}{\mathrm{GeV^{-1}}}\right)^2 \left( \frac{n_\mathrm{a}}{\mathrm{m^{-3}}}\right).
\end{align}
The upper graph of Fig.~\ref{fig:DRIFT_mod} shows the expected energy spectra of trapped solar KK axions at perihelion and aphelion.
The shape of \textcolor{black}{the} spectrum is assumed to be the same throughout the year, \textcolor{black}{and using the spectrum shown in \textcolor{black}{Fig.~\ref{fig:DRIFT_mod}} as input, the expected energy spectrum is calculated by a Monte Carlo simulation, which is shown in Fig~\ref{fig:signal_spectrum}. }

\begin{figure}\centering
\includegraphics[width=12cm]{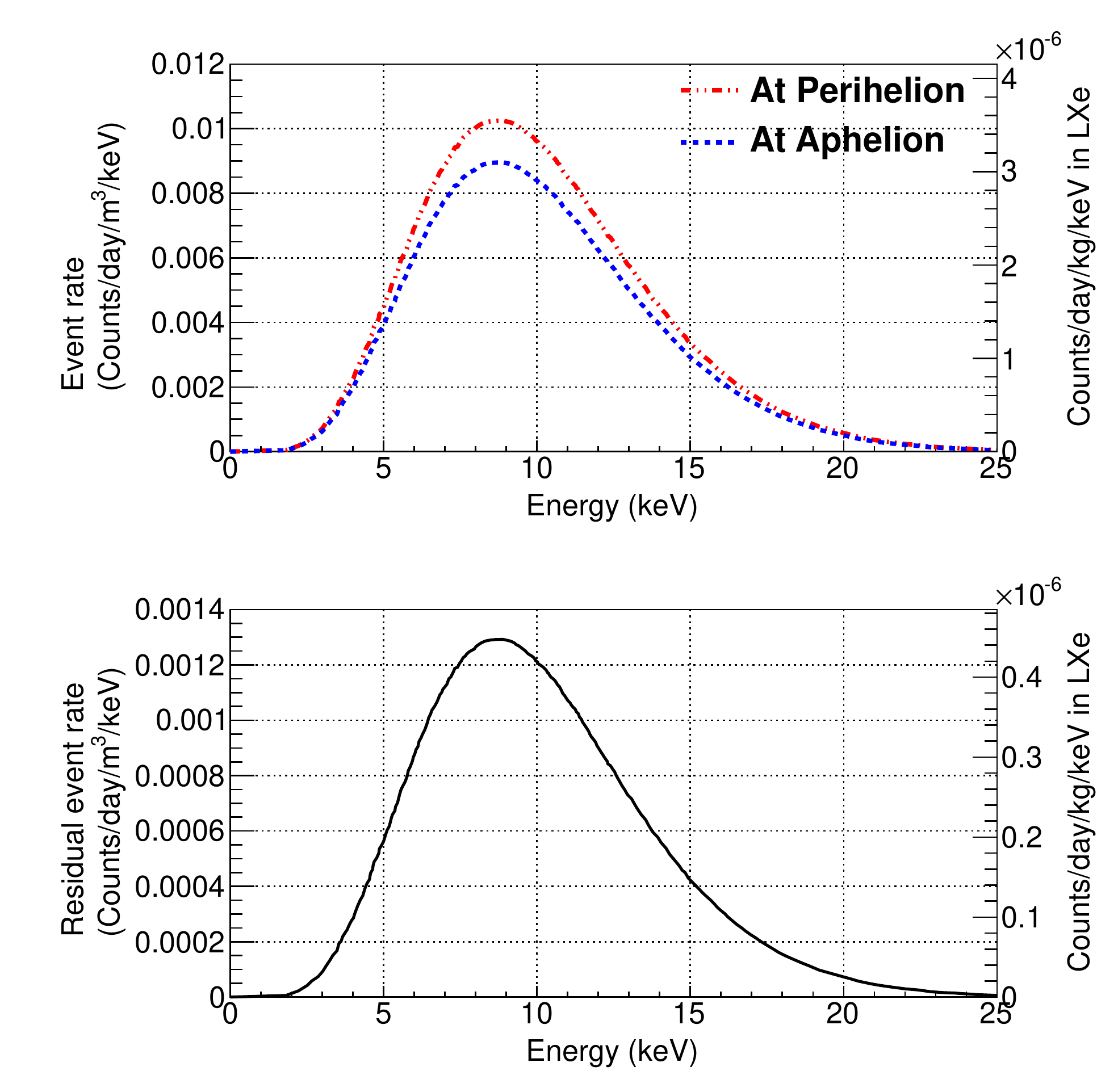}
\caption{Expected energy spectra of the sum of the two photons from the decay of trapped KK axions. The red dash-dotted curve and the blue dotted curve in the upper panel show the expected event rate at perihelion and aphelion, respectively. The black curve in the lower panel shows the expected residual event rate (difference between perihelion and aphelion). The energy spectra are taken from Ref. \cite{Morgan2005} and scaled according to the density at the Earth's position. A KK axion-photon coupling constant $g_{a\gamma\gamma} = 9.2 \times 10^{-14}\, \mathrm{GeV}^{-1}$ is assumed and KK axion number densities $n_\mathrm{a} = 4.36 \times 10^{13}\, \mathrm{m}^{-3}$ for perihelion and $n_\mathrm{a} = 3.81 \times 10^{13}\, \mathrm{m}^{-3}$ for aphelion are assumed. \textcolor{black}{The vertical axis on the right shows the expected event rate in liquid xenon.}}
\label{fig:DRIFT_mod}
\end{figure}
\begin{figure}\centering
\includegraphics[width=12cm]{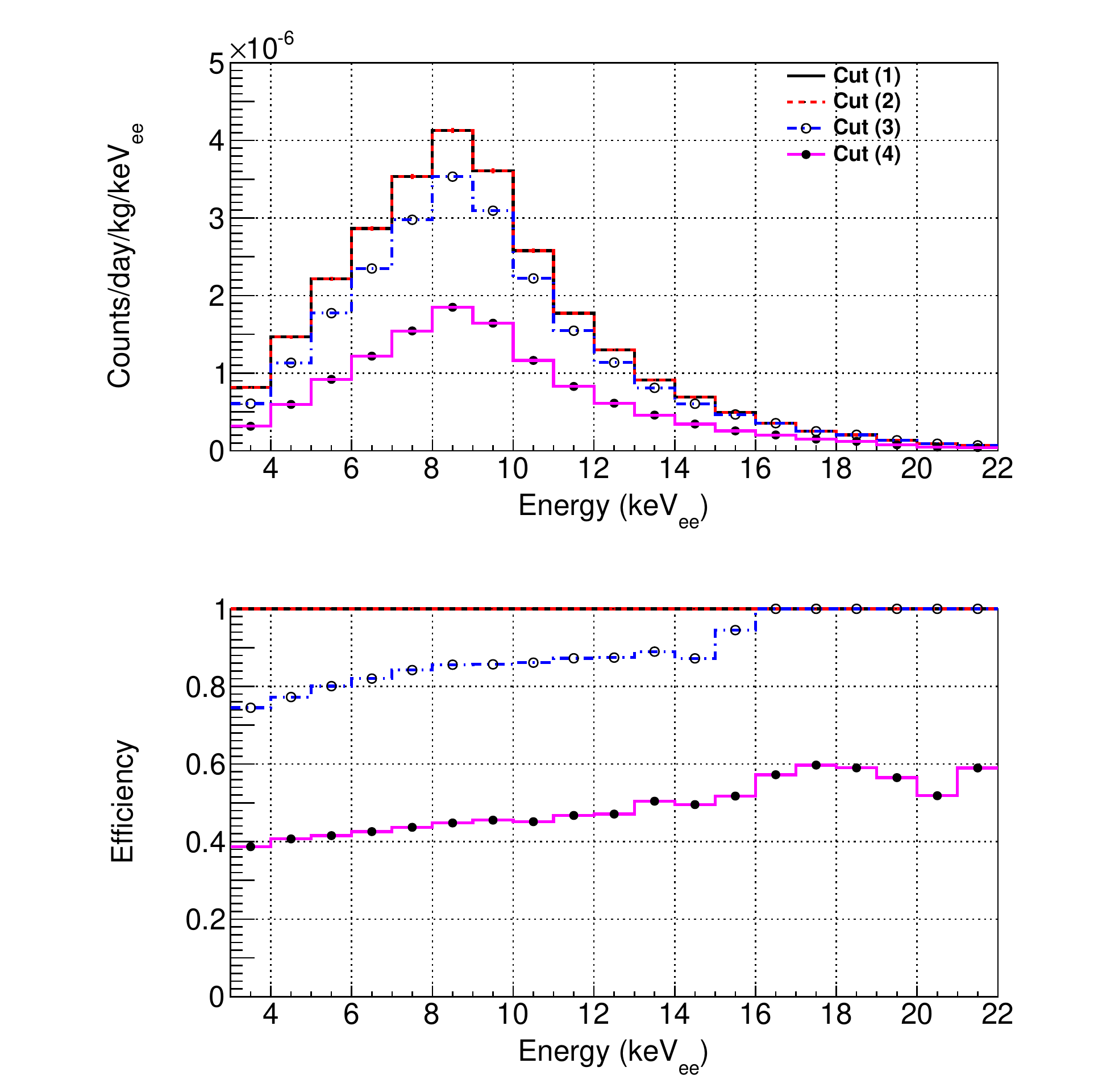}
\caption{Energy spectra of simulated KK axion events after the event selection steps (upper panel) and the corresponding selection efficiencies (lower panel). The histogram with the black solid line shows the events after cut (1). The red dotted line shows the events after cut(2), the blue dash-dotted line represents the events after cut (3), and the magenta solid line with filled circles shows the events after cut (4), which is the final sample. Here, $g_{a\gamma\gamma} = 9.2 \times 10^{-14}\, \mathrm{GeV}^{-1}$ and $n_\mathrm{a} = 4.07 \times 10^{13}\, \mathrm{m}^{-3}$  are assumed. \textcolor{black}{The difference between Fig.~\ref{fig:DRIFT_mod} and Fig.~\ref{fig:signal_spectrum} comes from the detector response and the non-linearity of the scintillation efficiency.}}
\centering \label{fig:signal_spectrum}
\end{figure}
%
%
%%%%%%%%%%%%%%%%%%%%%%%%%%%%%%%%%%%%%%%%%%%%%%%%%%%%%%%%%%%%%%%%%%%%%%%%%%%%%%
%                     section 4
%%%%%%%%%%%%%%%%%%%%%%%%%%%%%%%%%%%%%%%%%%%%%%%%%%%%%%%%%%%%%%%%%%%%%%%%%%%%%%
%
\section{Data analysis}\label{sec:analysis}
\begin{figure}\centering
\includegraphics[width=12cm]{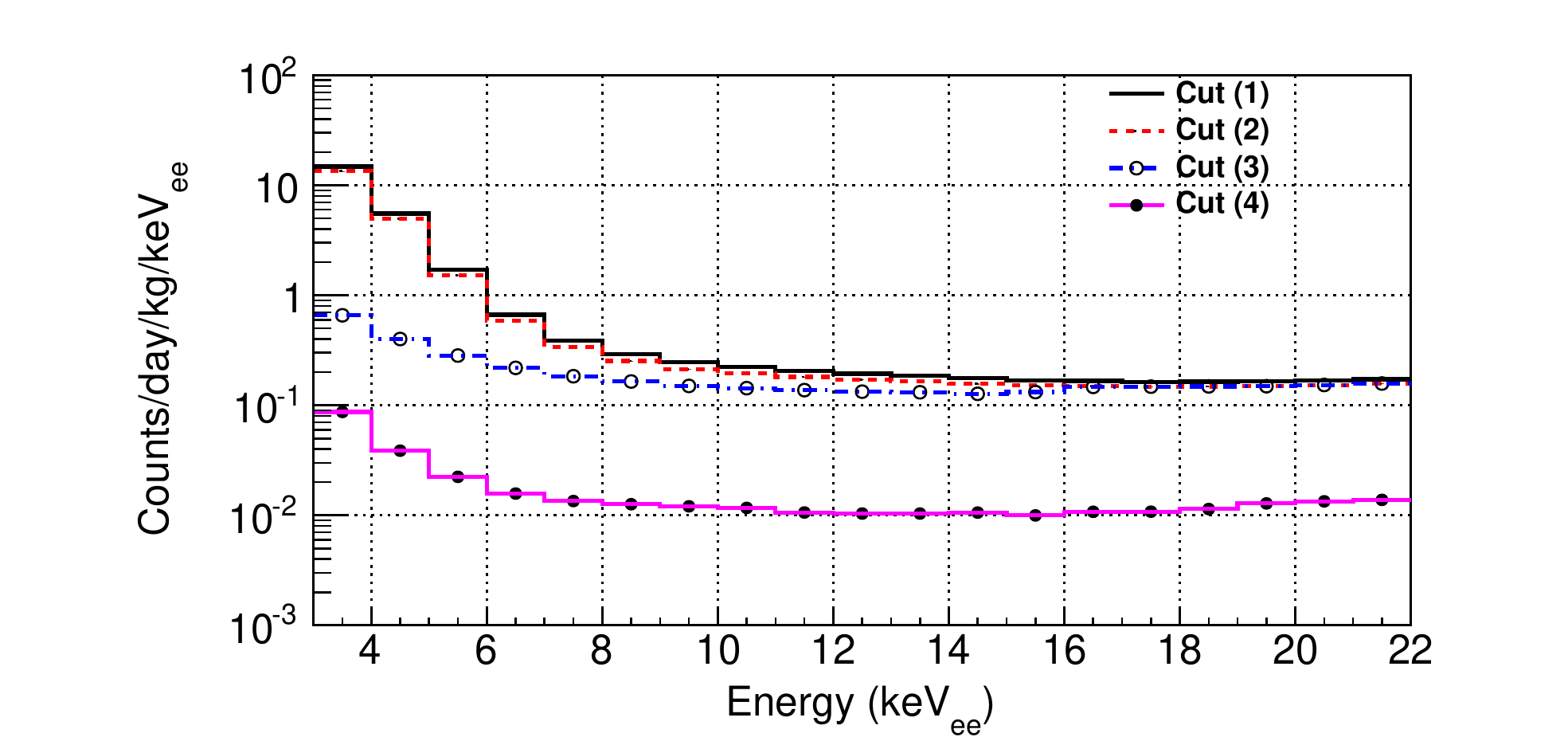}
\caption{Energy spectra of the observed data after each reduction step. The histogram with the black solid line shows the events after cut (1), the red dotted line shows the events after cut (2), the blue dash-dotted line represents the events after cut (3), and the magenta solid line with filled circle shows the events after cut (4), which is the final sample.}
\label{fig:data}
\end{figure}
Data from November 2013 to March 2015 with a total live time of 359 days is used for this study.
The data set is divided into 33 time-bins ($t_\mathrm{bins}$) with about 15 live-days each.
Four criteria are applied for event selection:
(1) The event is triggered only by the ID.
(2) The time difference from the previous event is more than 10 ms and the root mean square of hit timings in the event is less than 100 ns.
This cut is applied to reject events caused by afterpulses following bright events.
(3) The ratio of the number of hits in the first 20 ns to the total number of hits is less than 0.6.
This cut is applied only to the events with the total number of PE less than 200 in order to remove Cherenkov events originating from $^{40}$K decays in the photocathodes of the PMTs \cite{Detector}.
(4) The ratio of the largest number of PE detected by a single PMT to the total number of PE in the event has to be smaller than a certain value.
This cut is applied to remove background events occurring in front of a PMT window.
The cut value is a function of the total number of PE in the event and varies from about 0.2 at 8 PE to about 0.07 at 50 PE.
Details are described in \cite{Modulation}.
Figure~\ref{fig:data} shows the energy spectra after each selection step.
\textcolor{black}{Most of the remaining events originate from radioactivity in the aluminum seal of the PMTs \cite{Detector}.
Seasonal backgrounds, such as radon in the OD water and cosmogenics are found to be negligible.
}
\par
According to \textcolor{black}{our} weekly $^{57}$Co calibration, up to 10\% variation of the PE yield was observed throughout the measurement period.
This variation is understood to be due to the change of the liquid xenon absorption length which changed from about 4~m to 11~m \cite{Modulation}.
The cut efficiency is affected \textcolor{black}{by the variation of this PE yield, and changed up to 10\%.} 
We corrected this effect by accounting for relative cut efficiencies as evaluated from MC simulation.
The uncertainty of this cut efficiency is found to be the largest systematic error ($\sim \pm 5\%$ \textcolor{black}{on the event rate}), and enters as terms $K_{ij}$ into our $\chi ^2$ evaluation (see  Eq.~(\ref{eq:chi2})).
 Due to the normalization of  the overall cut efficiency, the systematic uncertainty of the relative cut efficiency becomes 0 at an absorption length of 8~m, and $K_{ij}$ becomes very small around March 2014.
The second largest contribution to the systematic error is the non-linearity of the scintillation yield ($\sim \pm 10\%$ \textcolor{black}{on the event rate}).
This uncertainty is taken into account in calculating the signal expectation, and is introduced as $L_{i}$ in Eq.~(\ref{eq:Rex}) below.
Between April and September 2014, a different way of calibrating electronics gains led to an additional 0.3\% uncertainty in the absolute energy scale of the experiment.
\textcolor{black}{This uncertainty is represented as $\sigma^2_{\mathrm{sys;}i,j}$ in Eq.~(\ref{eq:chi2}).}
Other systematic uncertainties were found to be negligible.
\par
The final data sample shown by the magenta spectrum in Fig.~\ref{fig:data} is evaluated for the KK axion search with an annual modulation method.
To this end an annual modulation amplitude is extracted from the data by a least Chi-squares fit.
The data in each time-bin are divided into \textcolor{black}{16} energy-bins ($E_\mathrm{bins}$) with a width of  $1\, \mathrm{keV_{ee}}$ each.
We used two pull terms \cite{Pull} of $\alpha$ and $\beta$ in the $\chi ^2$ which is defined as:
\begin{align}\label{eq:chi2}
\chi^2 = \sum _i ^{E_\mathrm{bins}}  \sum _j ^{t_\mathrm{bins}} \frac{\left(R^\mathrm{data}_{i,j} - R^\mathrm{ex}_{i,j} - \alpha K_{i,j} \right)^2}{\sigma^2_{\mathrm{stat};i,j} + \sigma^2_{\mathrm{sys;}i,j}}+ \alpha^2 + \beta^2,
\end{align}
where $R^\mathrm{data}_{i,j}$, $R^\mathrm{ex}_{i,j}$, $\sigma_{\mathrm{stat};i,j}$, $\sigma_{\mathrm{sys;}i,j}$ are the observed event rate, the expected event rate, and the statistical and uncorrelated systematic errors for each bin, respectively.
The subscripts $i$ and $j$ denote the respective energy and time bin.
$K_{i,j}$ represents the $1\sigma$ correlated systematic error based on the relative cut efficiency for each period.
$\alpha$ is the penalty term associated with $K_{i,j}$.
Based on Eq.~(\ref{eq:na}) and (\ref{eq:decay}), the expected event rate is defined as:
\begin{align}\label{eq:Rex}
R^\mathrm{ex}_{i,j} = \int ^{t_j+\frac{1}{2}\Delta t_j}_{t_j-\frac{1}{2}\Delta t_j} \left[ C_i + \xi\times (A_i  - \beta L_i)\left(\cos \frac{2\pi(t-t_0)}{T} + \frac{5}{2} e \cos^2 \frac{2\pi(t-t_0)}{T}\right) \right] dt,
\end{align}
where $\Delta t_j$ is the bin width of the $j$-th time bin. $C_i$ and $A_i$ are the constant term and the expected amplitude of the event rate in the $i$-th energy bin, respectively.
$A_{i}$ corresponds to half of the residual event rate in Fig.~\ref{fig:DRIFT_mod}.
\textcolor{black}{$L_i$, which is associated with the penalty term $\beta$, accounts for the uncertainty stemming from the non-linearity of the scintillation efficiency on $A_{i}$.}
$\xi$ is defined as $\xi = \frac{g_{\mathrm{a}\gamma\gamma}^{2}}{\left(9.2\times 10^{-14}\, \mathrm{GeV}^{-1}\right)^{2}} \frac{\bar{n}_\mathrm{a}}{4.07 \times 10^{13}\, \mathrm{m}^{-3}}$, and it represents the ratio of the expected amplitudes between the data and the considered model.
By treating $C_i$ and $\xi$ as free parameters in the fit, the $\chi^2$ is minimized.
The data is fit in the energy range between $3$ and $22\,\mathrm{keV_{ee}}$, excluding the range between $14$ and $17\,\mathrm{keV_{ee}}$.
This exclusion avoids systematic effects associated with the end of the range over which the Cherenkov cut is applied\textcolor{black}{, which can be seen in Fig.~\ref{fig:signal_spectrum}}.
%
%%%%%%%%%%%%%%%%%%%%%%%%%%%%%%%%%%%%%%%%%%%%%%%%%%%%%%%%%%%%%%%%%%%%%%%%%%%%%%
%                     section 5
%%%%%%%%%%%%%%%%%%%%%%%%%%%%%%%%%%%%%%%%%%%%%%%%%%%%%%%%%%%%%%%%%%%%%%%%%%%%%%
%
\section{Result and discussion}
\begin{figure}\centering
\includegraphics[width=15cm]{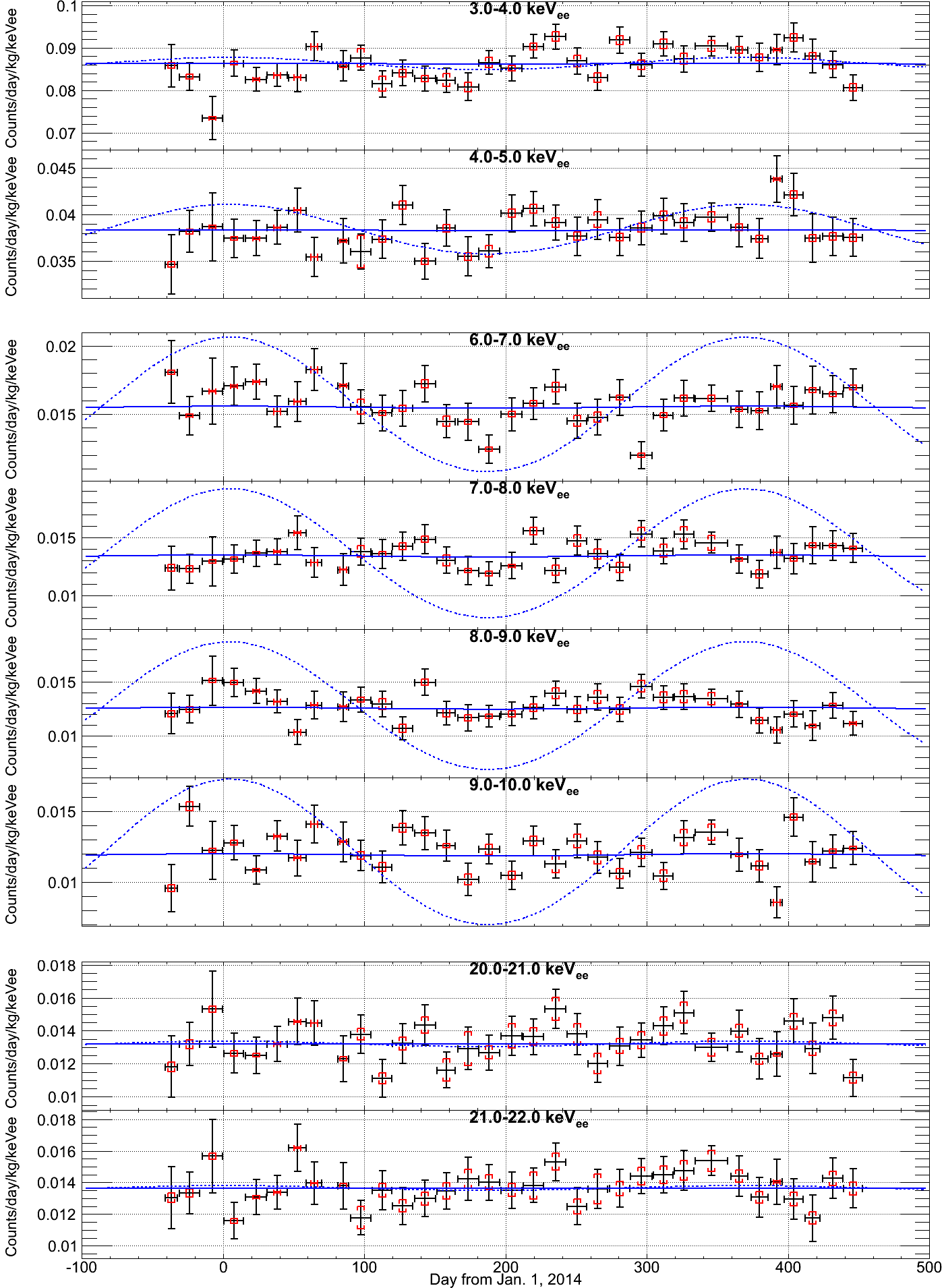}
\caption{Time variation of the observed event rate in representative energy bin. The horizontal axis is the time defined as the number of days from January 1st, 2014. The black points with error bars show the observed event rate for each period with statistical errors $\sigma_{\mathrm{stat};i,j}$. The red error bars show the systematic errors ($\sigma_{\mathrm{sys;}i,j}$ and $K_{i,j}$ are added in quadrature). The blue solid curves show the best fit result of the expected event rate variation ($\xi = 8.2 \times 10^{2}$). The blue dotted curves show the 20 times enhanced expected amplitudes of 90\% CL upper limit ($\xi = 2.7 \times 10^{3}$).}
\label{fig:timevar}
\end{figure}
Figure~\ref{fig:timevar} shows the event rate modulation and the best fit result for the expected event rate.
As a result, the fit obtains $\xi = 8.2 \times 10^{2}$ with $\chi ^2 / \mathrm{ndf} = 522.4 / 492$.
\textcolor{black}{We evaluated the significance of this result by using 10,000 no modulation dummy samples which have the same statistical and systematic errors as the data \cite{Modulation}.
This evaluation yields a p-value of 0.62.}
Since no significant excess in amplitude is found, a 90\% confidence level (CL) upper limit is set on the KK axion-photon coupling $g_{\mathrm{a}\gamma\gamma}$ as a function of the KK axion number density.
We use the likelihood ratio $\mathcal L$ defined as
\begin{align}
\mathcal L = \exp \left(- \frac{\chi^2 (\xi) - \chi^2_\mathrm{min}}{2}\right),
\end{align}
where $\chi^2 (\xi)$ is evaluated as a function $\xi$, while $\chi ^2_\mathrm{min}$ is the minimum $\chi ^2$ from the fit.
The 90\% CL upper limit is obtained by using the relation:
\begin{align}\label{eq:likelihood_ratio}
\frac{\int _0 ^{\xi_\mathrm{limit}}  \mathcal L\, d\xi}{\int _0 ^\infty \mathcal L\, d\xi} = 0.9.
\end{align}
The 90\% CL upper limit on the coupling constant derived for $\xi_\mathrm{limit} = 2.7 \times 10^{3}$ is
\begin{align}
g_{\mathrm{a}\gamma\gamma} < 4.8 \times 10^{-12}\, \mathrm{GeV}^{-1}\,\, (\mathrm{for}\,\, \bar{n}_\mathrm{a} =  4.07 \times 10^{13}\, \mathrm{m}^{-3}).
\end{align}
This limit  can be re-calculated for different KK axion densities and the obtained limit line is shown in Fig.~\ref{fig:limit}.
As a benchmark, the assumed solar KK axion model ($g_{\mathrm{a}\gamma\gamma} = 9.2 \times 10^{-14}\, \mathrm{GeV}^{-1},\, \bar{n}_\mathrm{a} =  4.07 \times 10^{13}\, \mathrm{m}^{-3}$) \cite{DiLella2003} is shown in Fig.~\ref{fig:limit}.
Note that the tension from the solar neutrino measurements as a consequence of the luminosity limit is  $L_\mathrm{a} < 0.1 L_\odot$ \cite{SolarNu} which corresponds to $\bar{n}_\mathrm{a} < 2 \times 10^{13} ~  \mathrm{m}^{-3}$, however, there still remains allowed parameter space for solar KK axion models with different values of $M_\mathrm{F}$ and $\delta$ as discussed in Ref.~\cite{DiLella2003}.
\begin{figure}\centering
\includegraphics[width=10cm]{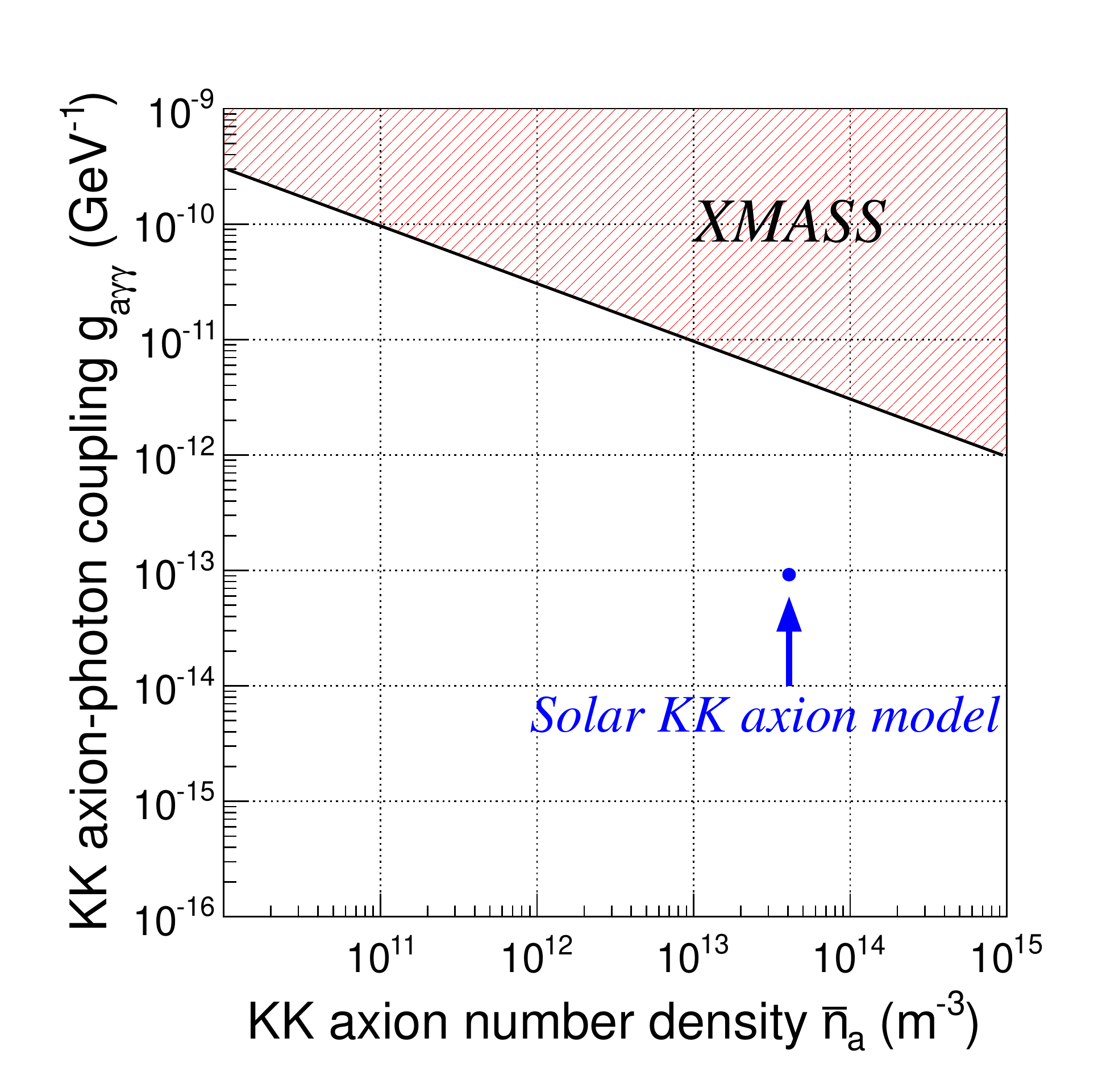}
\caption{The obtained 90\% CL excluded region from this work is shown by the black solid slope and the red hatched area. The model assumed in this study based on Ref.~\cite{DiLella2003} is indicated by the blue point. }
\label{fig:limit}
\end{figure}
\section{Conclusion}
We searched for the decay of solar KK axions by annual modulation using $832\times 359$ kg$\cdot$days of XMASS-I data.
No significant event rate modulation matched to the solar KK axion hypothesis ($n= \delta = 2$) is found, and a 90\% CL upper limit on the KK axion-photon coupling of $4.8 \times 10^{-12}\, \mathrm{GeV}^{-1}$ is obtained for $\bar{n}_\mathrm{a} = 4.07 \times 10^{13}\, \mathrm{m}^{-3}$. 
This is the first experimental constraint for KK axions.
\section*{Acknowledgments}
We gratefully acknowledge the cooperation of Kamioka Mining and Smelting Company.
This work was supported by the Japanese Ministry of Education, Culture, Sports, Science and Technology, Grant-in-Aid for Scientific Research (26104004, 26104005, 26104007), and N.~Oka thanks Grant-in-Aid for JSPS Research Fellow.
This work was also supported partially by the National Research Foundation of Korea Grant funded by the Korean Government (NRF-2011-220-C00006).
%
% can use a bibliography generated by BibTeX as a .bbl file
% BibTeX documentation can be easily obtained at:
% http://www.ctan.org/tex-archive/biblio/bibtex/contrib/doc/
%
%\bibliographystyle{ptephy}
%\bibliography{ptep}
%
% once the .bbl file has been generated then place the text in your article.

\end{document}